\documentclass[epj]{svjour}
\usepackage{graphics}
\usepackage{url}
\usepackage{tabularx}          
\usepackage[utf8]{inputenc}
\usepackage{array}
\usepackage{makecell}
\usepackage{booktabs}
\usepackage{multirow}
\usepackage{hyperref}
\usepackage{mathtools}
\usepackage{cite}
\usepackage{color}
\usepackage{colortbl}
\usepackage{ragged2e}
\begin{document}
\title{Numerical evaluation of a muon tomography system for imaging defects in concrete structures}
\author{Sridhar Tripathy\inst{1,2} Jaydeep Datta\inst{1,2} Nayana Majumdar\inst{1,2}\and Supratik Mukhopadhyay\inst{1,2}}% etc

\institute{Saha Institute of Nuclear Physics,\\AF Block, Sector 1, Salt Lake, Kolkata 700064, India \and Homi Bhabha National Institute, \\Training School Complex, Anushaktinagar, Mumbai 400094, India}
\date{Received: date / Revised version: date}
\abstract{Among numerous applications of muon tomography, deployment in civil structures has caught attraction of many recently. In this work, the appropriateness of muon scattering tomography to detect defects in concrete structures has been studied numerically. A few basic concrete structures that are frequently used in civil construction, have been considered as test cases. A simulation has been performed on Geant4 platform where an imaging setup built with several gaseous ionization detectors, having a specific spatial resolution for tracking the muons, have been modeled. The images of the test cases with and without the defect have been simulated for a month-long exposure of cosmic muons on the basis of their scattering from the composite concrete structures. The images have been compared using t-test to evaluate the performance of the imaging setup in identifying the defects. Further processing of the images has been done with a pattern recognition method proposed in our earlier work to improve defect identification. The efficacy of the said method has been evaluated in terms of the PRM-score devised in this work. The limitation and advantages of the present application of the muon scattering tomography encompassing the imaging and image processing technique in non-destructive evaluation of concrete structures have been discussed.} 
\PACS{
      {05.45.Pq}{numerical simulations}   \and
      {29.40.Cs}{Ionization chambers}
 %    } % end of PACS codes
} %end of abstract
\maketitle
\section{Introduction}
\label{intro}
\label{sec:intro}
Cosmic ray muons are high-energy charged particles with deep penetration power. Their invasive nature and omnipresent property inspire to utilize the scattering of the muons for imaging applications. Muon Scattering Tomography (MST) works on the basis of deviation of cosmic muons from their path due to their interaction with atomic nuclei and electrons of the target material. These particles with their large rest mass ($\approx$105 MeV/c$^{2}$) and momentum (mean $\approx$ 4 GeV/c) pass through the whole atmosphere with minor deviations~\cite{pdg}. A large section of muons is even capable of traversing through rocks, civil structures, etc. The deviation and energy loss of muons occur due to physical processes like multiple Coulomb scattering (mcs) and ionization. The mcs angle distribution is considered as a Gaussian distribution~\cite{Highland1975,Lynch1991} with width $\sigma$ given by equation~\ref{eq:scatt}. 

\begin{equation}
\label{eq:scatt}
\sigma = {13.6 \over \beta p}{\sqrt {L \over X_0}}\Big(1 + 0.038 ln{L \over X_0}\Big)
\end{equation}
\begin{equation}
\label{eq:rL}
X_0 = {716.4 A \over {\rho Z(Z+1) ln\Big({287 \over {\sqrt Z}}\Big)}}
\end{equation}
Here, $p$ is the momentum of muon, $\beta=v/c$ is the ratio of the speed of muon, $v$, to that of light, $c$. The term, $L/X_{0}$, is the thickness of the scattering medium in terms of its radiation length, $X_{0}$. As shown in equation~\ref{eq:rL}, $X_{0}$ is related to the atomic weight, $A$, atomic number $Z$, and density, $\rho$ of the medium. The equations~\ref{eq:scatt} and~\ref{eq:rL} suggest that the magnitude of deviation of muons significantly depends on $Z$ and $\rho$. Therefore, while traversing through high-Z and dense matter, such as, lead, uranium, muons undergo larger deviations. On the other hand, low-Z and lighter materials, like concrete, aluminium, can only cause feeble deviation in their tracks. We have shown in our previous work~\cite{Tripathy_2020} that on the basis of scattering property of muons, high-Z, mid-Z, and low-Z materials can be distinguished convincingly with the given imaging setup and imaging processing technique based on a Pattern Recognition Method (PRM).

The cosmic ray muon radiography has been established as a non-invasive imaging technique since last couple of decades. It has been considered for many crucial applications, such as, identifying fissile materials~\cite{Miyadera2013,LaRocca2014}, scanning cargo containers~\cite{Borozdin2003,Morris2008,Riggi2013,Thomay2013,Harel2019}, monitoring nuclear waste containers~\cite{Jonkmans2013a,Chatzidakis2016b}, investigation inside geological structures~\cite{Nagamine1995,Marteau2012,Morishima2017,Bonechi2020}, water towers~\cite{Bouteille2016} etc. It has been observed that for carrying out imaging of large concrete structures, muon transmission radiography is the preferred option~\cite{Bonechi2020,Alvarez1970,Morishima2017} by which images of the targets are produced on the basis of absorption of muons inside the target. For discriminating or imaging high-Z materials, MST has been used conveniently as the scattering angle becomes sufficiently large with increase in $Z$ and $\rho$. Apart from imaging based on scattering and stopping of muons, there has been example of devising a new method for imaging very low-Z and extremely light material using muon-induced secondary radiations. A unique application of the said process in imaging of organic soft tissue, polymethyl methacrylate and water can be found in \cite{Mrdja2016} which would be impractical to achieve using MST.  

The imaging of civil structures has turned out a necessity of modern age to maintain the civil works, like, buildings, bridges, highways, tunnels, dams, etc. Several standard Non-Destructive Evaluation (NDE) techniques to mention are, namely, ultrasonic tomography~\cite{Lorenzi2016,Dong2016}, infrared (IR) tomography~\cite{Abdel-Qader2008,Maierhofer2010}, Ground Penetrating Radar (GPR)~\cite{Bungey2004}, impact-echo~\cite{Sansalone1997}, etc., which have been practiced widely for many decades to address diverse issues of imaging. For example, the ultrasound technology has been found very fruitful in identifying delamination of concrete bridge decks, corrosion of steel rebars~\cite{DiBenedetti2013}, but it turns out difficult for ultrasonic waves to penetrate attenuating materials, like, steel and fibre reinforced plastic~\cite{Feng2002} which are used as outer layers of composite elements in civil structures. Similarly, IR thermography is successful in scanning bridges, roads and airport pavements~\cite{Poston1995,Avdelidis2003}. However, there are some limitations as well, such as, requirement of a large amount of heat to image non-conducting materials (active thermography), interference of environmental parameters, like, sun-light intensity (passive thermography), shadow of other bodies, wind etc~\cite{Abdel-Qader2008}. The radiological imaging methods using X-ray and gamma-ray are very effective but not preferred due to their potential of biological hazards. On the contrary, the minimum-interacting property leading to high-penetration and biologically non-hazardous nature of cosmic muons along with their abundant availability gives the muon transmission or scattering tomography an edge over other imaging techniques. In~\cite{Dobrowolska2020}, MST has been advocated to be used as a NDE technique for its application in civil imaging with critical comparison to others.

Although the application of MST for imaging concrete-like light materials can be non-trivial, there have been several applications of imaging iron bars inside Reinforced Cement Concrete (RCC) structures~\cite{Dobrowolska2020,Guardincerri2016,Niederleithinger2020a}. The promising outcomes of these investigations are encouraging for using MST in imaging concrete structures. In the present work, an attempt has been made to go one step beyond and investigate the credibility of MST in identifying defects in concrete structures using images produced by a given MST setup. A few cases of defects, such as, air voids, corrosion of metallic components, etc., that appear frequently in civil structures, have been opted as test cases in this work. Each of these problems is unique in nature and possesses different kinds of challenges for imaging. These defective structures have been imaged using a MST setup, consisting of gaseous ionization detectors for tracking the muons, in Geant4~\cite{Agostinelli2003} simulation framework. The flexibility in design, low-cost fabrication along with suitable spatial and temporal resolutions, make the gaseous ionization detectors one of the most suitable options to be used for muon tomography applications~\cite{Baesso2014,Bouteille2016}. The simulated images have been further processed and analysed with the t-statistics and PRM to study efficacy and limitation of the present method of MST for its application in imaging of civil works.

The content of this article has been organised in the following way. Details of the imaging setup, detector dimensions, and their placement, target specification, etc., have been described in section~\ref{sec:2}. The algorithm for defect identification and evaluation of degree of discrimination has been narrated in section~\ref{sec:3}. The imaging results and their analysis can be found in section~\ref{sec:4} and the conclusive statements in section~\ref{sec:5}.  

\section{Simulation scenarios}
\label{sec:2}
\begin{figure}[!htbp]
\centering
\resizebox{0.75\textwidth}{!}{%
 \includegraphics{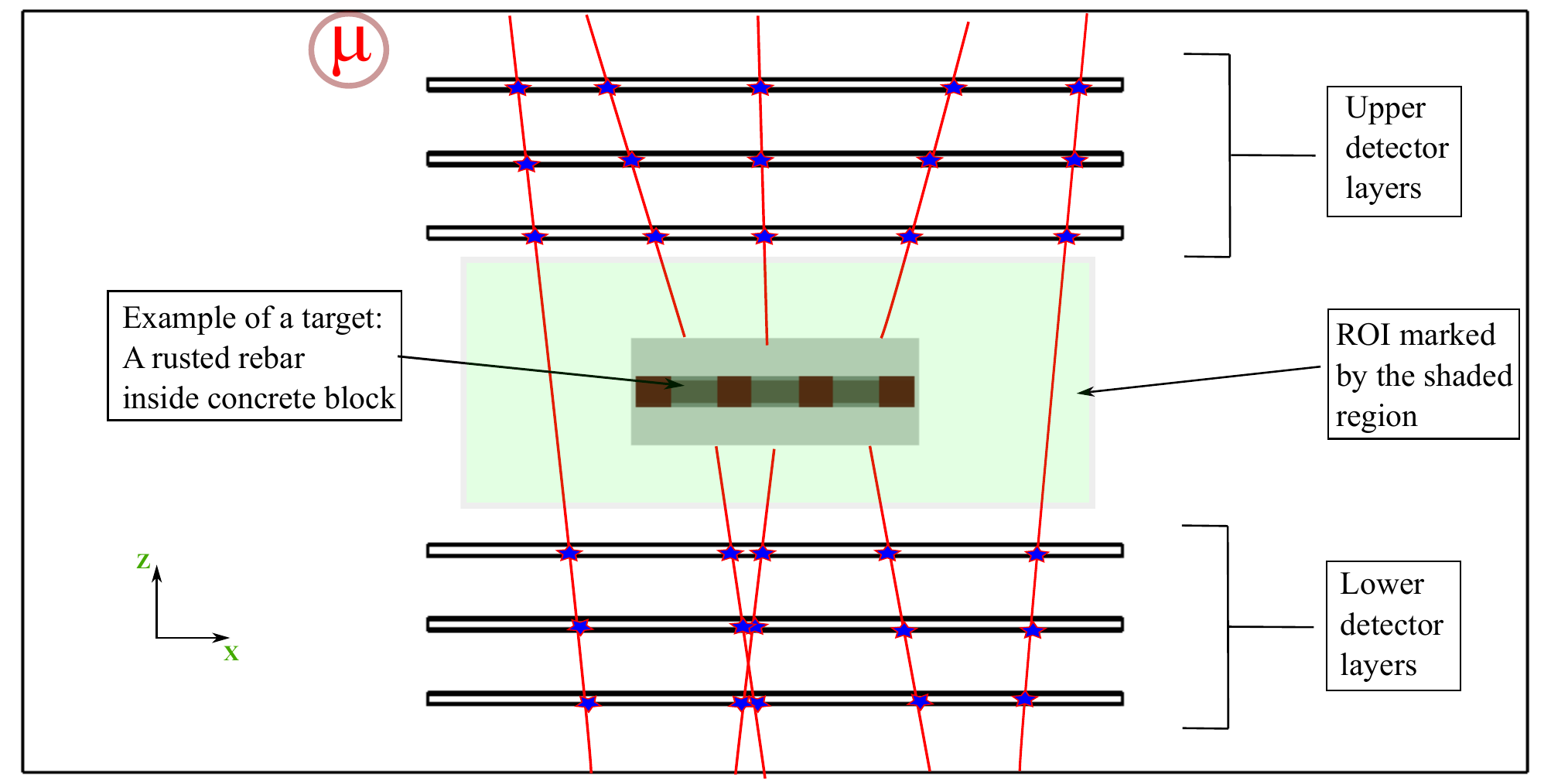}
}
\caption{MST imaging setup as modelled in Geant4. The muon-hits on the tracking detectors have been marked with stars.}
\label{fig:sch}       % Give a unique label
\end{figure}
The MST setup constructed in Geant4 for imaging several test cases have been shown in figure~\ref{fig:sch} along with an instance of a target. Two sets of three tracking detectors have been placed in parallel with a separation of 7 cm on either sides of the Region Of Interest (ROI) where the targets are to be placed (marked with light green colour). The detectors, each having a gas thickness of 2 mm have acted as trackers in order to record two-dimensional (2D) position information of the muon-hits (marked in red star). A spatial resolution of 200 $\mu$m of the detectors has been implemented by introducing a random uncertainty to the $X$ and $Y$-coordinates of the respective muon-hits. The area of the detectors govern the solid angle acceptance of the setup and hence has to be optimized for uniform exposure across the ROI. Thus, for each test case, it has been varied according to the target size as mentioned in table~\ref{set}. The muon events have been generated using Cosmic Ray Library (CRY)~\cite{Hagmann2007}. A cosmic ray exposure equivalent to 30 days has been used for imaging the test cases. 

\begin{table}[!htbp]
	\centering	
\begin{tabular}{|p{2cm}|p{3.5cm}|p{3.5cm}|p{3.7cm}|}
\hline
\textbf{Specification} & \textbf{Rusted Rebar} & \textbf{Void in} & \textbf{Void in Concrete Deck}\\
& &\textbf{CFST}&\textbf{Deck}\\
\hline
Detector Area & 60$\times$60 cm$^2$ & 100$\times$100 cm$^2$ & 140$\times$140 cm$^2$\\
\hline
Target & \textbf{Concrete Volume:} 25$\times$10$\times$10 cm$^{3}$  \newline
\textbf{Rebar Length:} 24 cm \newline
\textbf{Rebar Dia:} 3 cm & \textbf{CFST Length:} 30 cm \newline
\textbf{CFST Dia:} 16 cm \newline
\textbf{Steel Thickness:} 5 mm &\textbf{Concrete Volume:} 80$\times$80$\times$15 cm$^{3}$ \\
\hline
Defect & \textbf{Rust Thickness:} 4.5, 2.25 mm & \textbf{Void Thickness:} 10, 7 mm & \textbf{Spherical Void Dia:} 6.74, 5.64 cm \newline
\textbf{Void Cubical Void Side:} 6, 5 cm\\
\hline
Defect Percentage along $Z$ & 30\% , 15\% & 12.5\%, 8.75\% & 40\%, 33.33\%\\
\hline
\end{tabular}
	\caption{\label{set} The specifications of the targets and the detectors as used in GEANT4 simulation of three test cases. Here dia. stands for diameter.}
\end{table}

Three examples of typical defects commonly found in in concrete structures~\cite{Abdel-Qader2008,Oshita2015,Dong2016}, have been considered as test cases in the present work. For each case, a variation of defect dimension has been simulated to validate consistency of the imaging technique and study its limitation as well. The percentage of defects along the $Z$-direction (the direction of cosmic muon exposure) has been furnished for all the cases in table~\ref{set}. A brief description of each of the test cases has been provided in the following sub-sections.
\subsection{Rusted rebar}
\begin{figure}[!htbp]
\centering
\resizebox{0.8\textwidth}{!}{%
 \includegraphics{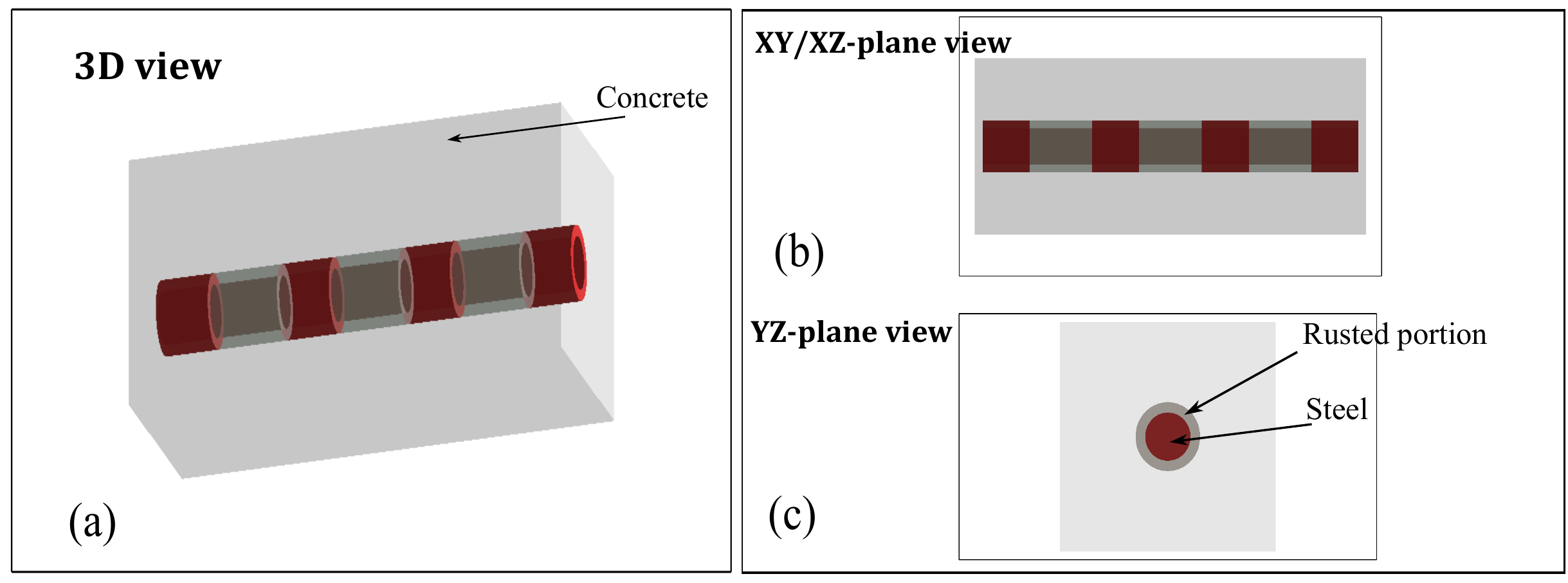}
}
\caption{Three different views of the geometry of the rusted rebar as constructed in Geant4. The concrete block (light grey), steel rebar (red) and rust (dark grey) have been marked. The target is placed at the center of the ROI with its central axis lying along the $X$-direction.}
\label{JoinRebar}       % Give a unique label
\end{figure}

RCC structures with steel rebars have been used for a long time to build civil structures. The most common problem in such structures is the rusting followed by corrosion of rebars which is caused by their exposure to the atmosphere, rainfall, concrete-metal contacts, etc~\cite{sourav}. Various NDE techniques have been implemented to identify the rusted region in rebars, such as, half-potential method~\cite{halfpotential}, thermal imaging~\cite{Oshita2015,Na2019}. In~\cite{Oshita2015}, a quantitative measurement of corrosion has been done by thermography using electromagnetic induction. A  similar structure but with multiple corroded regions has been constructed in Geant4 in the present work. Three different views of the geometry have been shown in figure~\ref{JoinRebar} where all the materials, namely, the concrete (in light grey), steel (in red) and rust (in dark grey) have been marked. The RCC volume with the steel rebar embedded at its centre has been placed at the middle of the ROI with its central axis lying along the $X$-direction. In the simulation model, Fe$_{2}$O$_{3}$ composition has been used as rust with density 5.25 g/cc while steel and concrete have been simulated with densities 7.87 g/cc and 2.3 g/cc, respectively. The thickness of the rusted-ring has been varied following two different values as mentioned in table~\ref{set}.

\subsection{Void in CFST}
\begin{figure}[!htbp]
\centering
\resizebox{0.8\textwidth}{!}{%
 \includegraphics{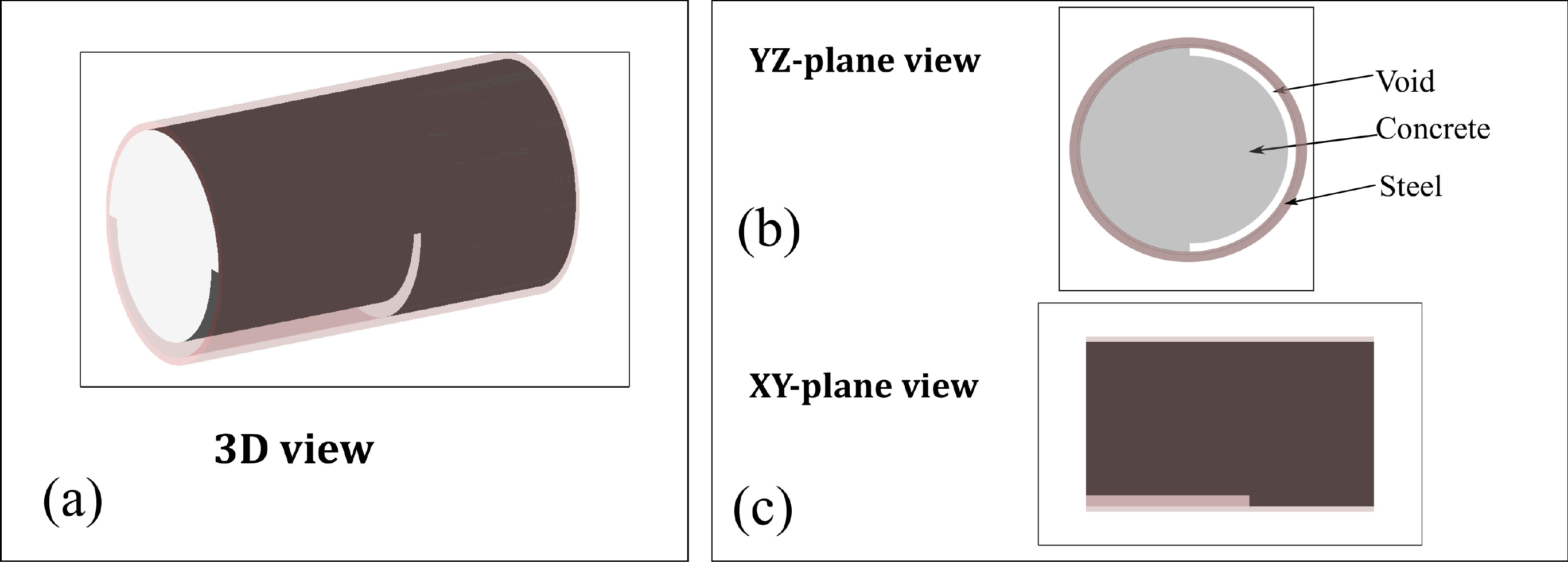}
}
\caption{Three different views of the geometry of the defective CFST as considered in Geant4. The steel tube (in light brown), concrete (in light grey) and void (in white) have been marked in $YZ$-plane view. The target has been positioned at the center of the ROI with its central axis lying along the $X$-direction.}
\label{JoinCFST}       % Give a unique label
\end{figure}

Concrete-Filled Steel Tube (CFST) is a cost-effective solutions for implementing in large numbers in constructing truss elements and columns in high-rise buildings. The comprehensive strength of concrete and confinement along with the rigidity of steel can make the combination carry more load than individual elements~\cite{Shanmugam2001}. However, voids, de-bonding ring-gaps can occur between concrete and steel due to fluidity before initial setting or dry shrinking during use~\cite{Xue2012}. These types of defects reduce the load carrying capacity of the CFST which is considered as a critical problem in civil engineering. Due to shielding effect, high density and low radiation length of steel, imaging of defects of CFSTs is a difficult task for traditional NDE techniques such as, electromagnetic waves, impact echo technique, X-ray, and gamma ray~\cite{Chen2019}. Several experimental works have been done to identify and image the de-bonding occurring in CFSTs~\cite{Dong2016,Xu2013,Chen2020a}. Similar to one of the specimen structures described in~\cite{Dong2016}, a test case has been constructed in Geant4 with 50\% circumferential void near the steel edge. Three different views of the geometry have been shown in figure~\ref{JoinCFST} where the steel tube (light brown), concrete (in light grey) and void (white) are marked in $YZ$-plane view. The CFST has been placed at the center of the ROI with its central axis lying in the $X$-direction. Unlike, the previous example of defective rebar, the CFST defect is not symmetric to the rotation in $YZ$-plane. Therefore, to allow maximum exposure of the muons to the void region, the CFST is placed such that it faces one side ($Y$-direction), as shown $YZ$-plane view. The thickness of the circumferential void has been varied as mentioned in table~\ref{set}.

\subsection{Voids in concrete deck}
\label{sec:deck}
\begin{figure}[!htbp]
\centering
\resizebox{0.75\textwidth}{!}{%
 \includegraphics{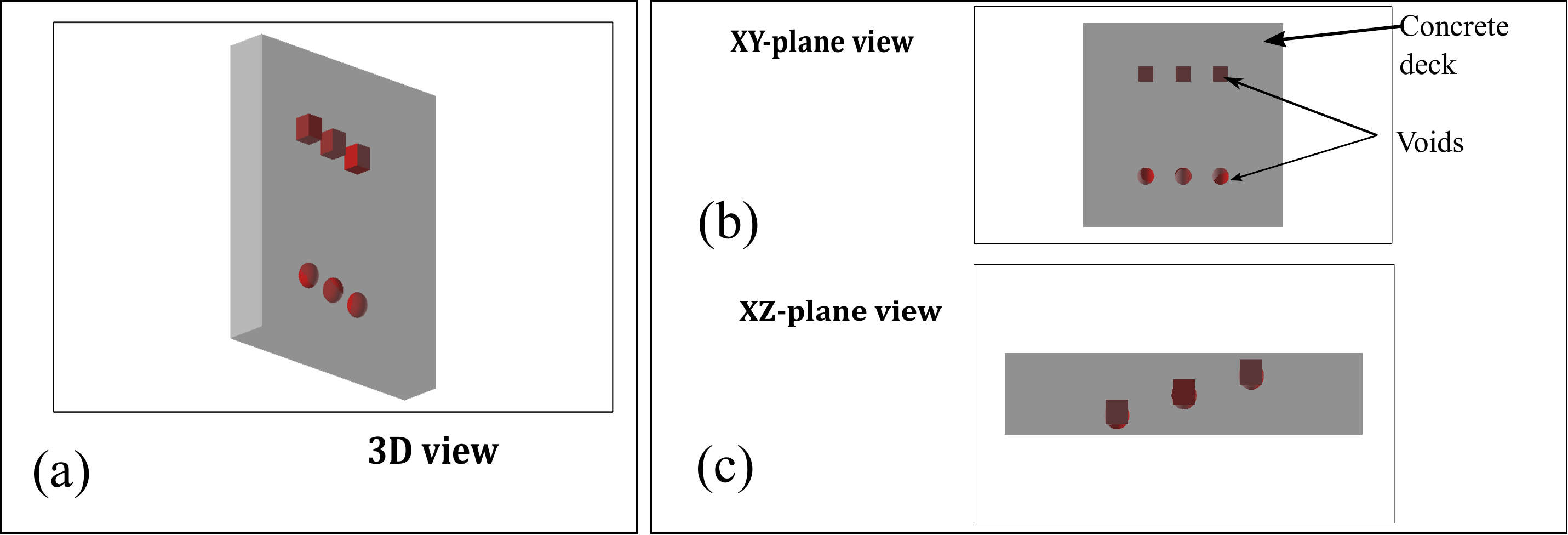}
}
\caption{Three different views of the geometry of the defective concrete deck as considered in Geant4. The concrete deck (in dark grey) and spherical and cubical voids (in red) have been marked. The deck has been positioned in the $XY$-plane at the center of ROI. The placement of the voids at different depths (4, 8 and 12 cm from the top) has been shown in the $XZ$-plane view.}
\label{JoinVoid}       % Give a unique label
\end{figure}

Another common problem found in concrete structures is sub-surface voids and delaminations in concrete decks. These defects make the bridge decks structurally deficient. Different NDE techniques, such as, IR thermography, GPR, etc., have been used to identify the delaminations and voids in concrete decks~\cite{Maierhofer2006,Tarussov2013}. The detection of these sub-surface defects depends upon their size and position inside the concrete deck. In some research works carried out to identify such defect~\cite{Maierhofer2010,Abdel-Qader2008}, it has been reported that the amount of concrete covering has a substantial impact on imaging the defects. Therefore, it would be easier to image defects if they are not buried deep inside. In~\cite{Abdel-Qader2008}, IR imaging has been used to detect shallow defects inside concrete decks where a few bridge decks have been constructed with several voids and delaminations. In this work, a similar kind of geometry has been constructed in Geant4 with some voids. Three different views of the geometry have been shown in figure~\ref{JoinVoid} where the conrete deck (in dark grey) and voids (red) have been marked in the $XY$-plane view. The voids in each row have been placed in three different depths (4, 8 and 12 cm from the top) to consider the effect of randomness in concrete cover. In reality, however, defects are not expected to follow any such defined pattern. In order to maintain arbitrariness in the nature of the voids, two different shapes of voids (spherical and cubical) have been implemented. The cross-section of the voids have been made comparable to receive similar muon exposure, \textit{i.e.}, for length of the cubical void, $a$ and the diameter, $d$ of the spherical void, their cross-section remains the same $(\pi d^{2}/4=a^{2})$. Two different dimensions of the voids have been used as mentioned in table~\ref{set}.

\section{Detection methodology}
\label{sec:3}
In MST, the incoming and outgoing tracks of cosmic muons are used to find scattering vertices in the ROI and determine the scattering angle ($\theta$). In literature, various algorithms can be found to identify the scattering vertices~\cite{Schultz2004,Thomay2013,Riggi2013,Chatzidakis2018}. The Point of Closest Approach (PoCA)~\cite{poca,Tripathy2018,Schultz2004} is a reasonably accurate and relatively simpler algorithm among them which determines a single scattering vertex from the intersection or the closest approach of the extrapolated incoming and outgoing tracks. Although the mcs process is not considered in this algorithm, it is widely used due to its fast computation and fairly precise performance~\cite{Yang2019,Bonechi2020}. In the present work, the PoCA algorithm has been implemented to determine scattering vertices and scattering angle using C++ with links to ROOT~\cite{Brun1996} for data handling and plotting.

\begin{figure}[!htbp]
\centering
\resizebox{\textwidth}{!}{%
 \includegraphics{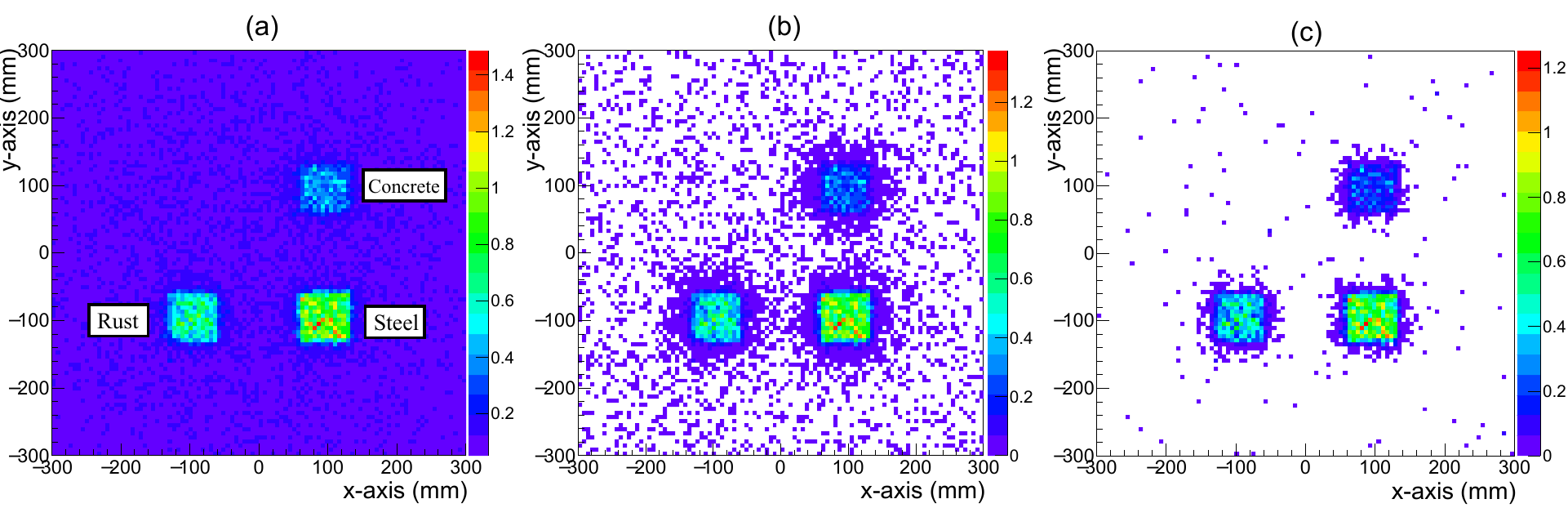}
}
\caption{Images obtained for blocks of steel, concrete, rust for (a) no threshold, (b) $\theta_{th}$=5 mrad, (c) $\theta_{th}$=10 mrad. The images have been obtained for half-an-hour of cosmic muon exposure.}
\label{fig:th}       % Give a unique label
\end{figure}

The vertices obtained with small scattering make it difficult to distinguish targets from the empty space filled with air which we would refer as background. Hence, it is necessary to use a threshold of the scattering angle, $\theta_{th}$, to filter the background which actually depends on the target to be identified. In context of the present work, a simulation has been performed with three cubical targets each of side 7 cm, constructed from steel, concrete and rust which are commonly found in any civil structure. Figure~\ref{fig:th} shows the images obtained on the basis of projected scattering vertices for a muon exposure of half-an-hour following the method described in~\cite{Tripathy_2020} and further filtered with three different $\theta_{th}$ values. It is evident from the images that the target-to-background (signal-to-noise) ratio improves with the increase in $\theta_{th}$. As a result, the threshold, $\theta_{th}$ = 10 mrad has been opted for the present numerical work to filter the background or the noise. The scattering vertices have been then projected to the central $XY$-plane of the ROI to obtain a 2D map. It has been further pixeled with a definite size and each pixel has been weighted by a parameter $S$ as defined in equation~\ref{eq:s}.
\begin{equation}
\label{eq:s}
S=\sum_{k=1}^{\rho_{c}}{\theta_{k}}
\end{equation}

\justify 
where $\rho_{c}$ represents the number scattering vertices in the pixel and $\theta_{k}$ is the scattering angle at the $k$th vertex. It can be followed from the equation~\ref{eq:s} that $S$-parameter is the sum of the scattering angles at all scattering vertices in the given pixel that passed through the filter. This weighted 2D scattering map (we would refer to as $S$-map henceforth) has been subjected to further analysis. The discrimination of the defective targets from the perfect targets (without any defect) has been done following a statistical test and the PRM technique applied on the $S$-maps obtained for different test cases. 

\subsection{Discrimination significance by t-statistics}
The t-statistics is a widely accepted analysis tool for comparing the mean of a test distribution to that of a reference to determine if there is a significant difference between them. It is quantified by the t-value which measures the difference relative to the variation of the reference distribution. The larger the magnitude of t-value the greater is the departure from the null hypothesis that considers both the distributions to be identical ~\cite{cowan}. In this present work, the t-test has been used for comparing the $S$-maps of the target with and without the defect in each of the test cases. It has been calculated in each case using equation~\ref{t1}. 

\begin{equation}
\label{t1}
\begin{split}
    t=\frac{\mu_{1}-\mu_{2}}{s_{v}\left[\frac{1}{n_{1}}+\frac{1}{n_{2}}\right]}\\
    \textrm{with}\;\;\;
    s_{v}=\sqrt{\left[\frac{(n_{1}-1){s_{1}}^2+(n_{2}-1){s_{2}}^2}{n_{1}+n_{2}-2}\right]}
\end{split}
\end{equation}

\justify
where $\mu_{1}$, $n_{1}$ and $s_{1}$ are the  mean, size and standard deviation, respectively,  of the reference distribution ($S$-map of the perfect target) while $\mu_{2}$, $n_{2}$ and $s_{2}$ are the corresponding parameters of the test distribution ($S$-map of the defective target). The $p$-value, which is the probability of finding the observed t-value or more extreme given the null hypothesis is true (the $S$-maps of the perfect and defective targets are identical), has been estimated. From the $p$-value the statistical significance of the discrimination has been determined. The significance of 2$\sigma$ has been considered as the threshold for identifying the defects in test data. 

\subsection {Discrimination capability using PRM-score}
\label{sec:PRM}
\begin{figure}
\centering
\resizebox{0.75\textwidth}{!}{%
 \includegraphics{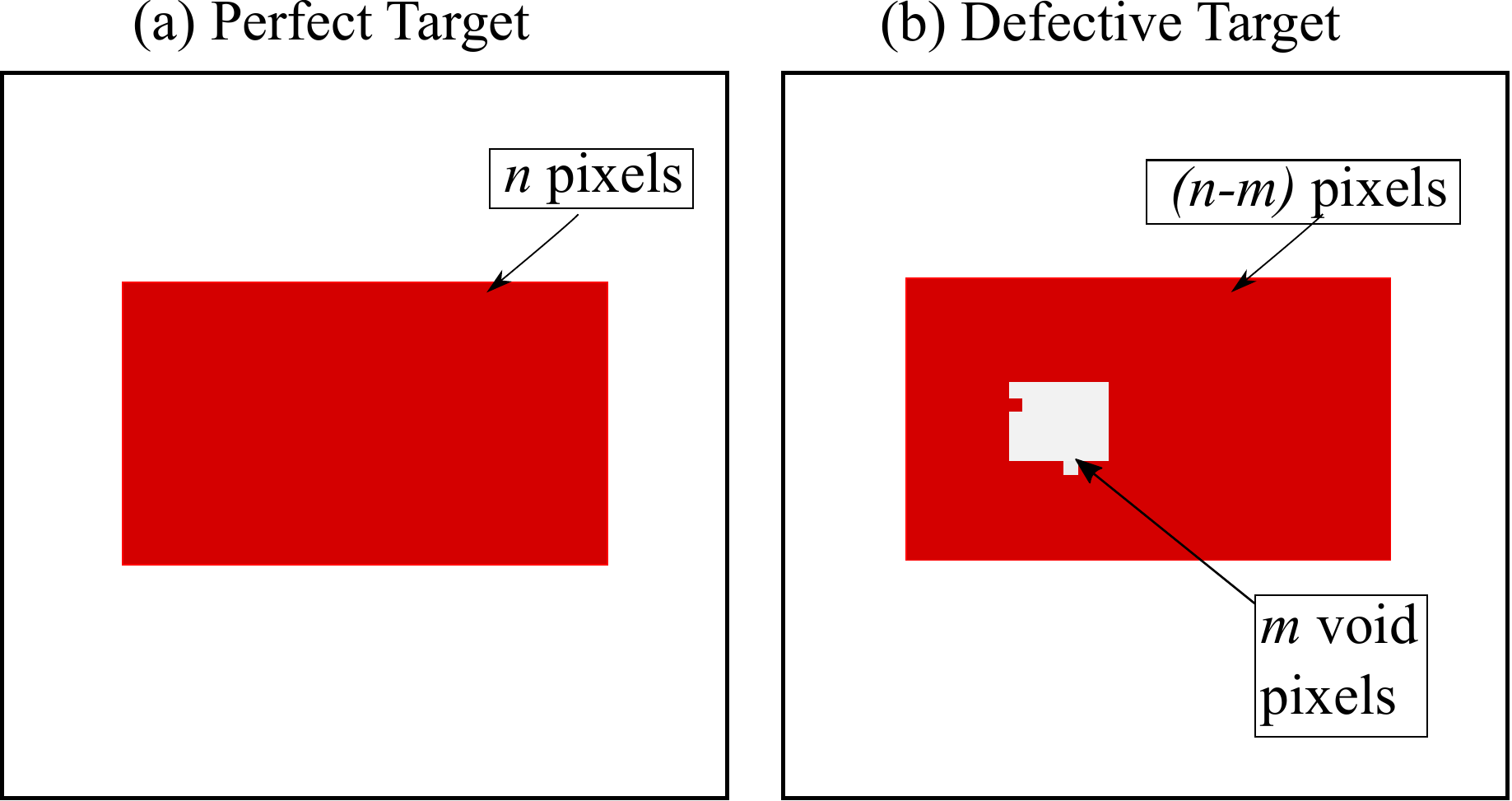}
}
\caption{Comparison of images for perfect target (reference) and defective target (test) based on PRM-score.}
\label{fig:prm}       % Give a unique label
\end{figure}

In~\cite{Tripathy_2020}, the basics of processing images with PRM method have been explained in detail. To follow the procedure, the $S$-map of the ROI holding the target has been converted to a numerical matrix `$R$'. From the reference $S$-map of the perfect target (say $R_{p}$) a filter sub-matrix `$K$' has been obtained which represents the physical properties ($Z$ and $\rho$) of the reference in terms of the $S$-parameter. The sub-matrix `$K$' has been subsequently convoluted with the `$R$' matrix of the test target (say $R_{d}$) to probe for similarity. The regions of the ROI $S$-map which have turned out smaller in terms of $S$-parameter than that of the filter sub-matrix, have been rejected. The rest areas which are similar or larger with the filtering $S$-parameter have been passed. The technique, seemingly equivalent to a high-pass filter has been found capable to identify multiple targets in the same ROI perform reasonably well in complex scenarios such as, when target is submerged inside background,different shapes of the target, etc. In the present work, to numerically quantify the degree of discrimination a metric, namely, PRM-score has been introduced which evaluates the similarity between the reference and test targets as well as the capability of discriminating them. To put it in a simple way, the PRM-score specifies the difference between images reconstructed after PRM-processing of the test and the reference target measured in the units of $\delta n$, which is the random error arsing out of repeated measurement of PRM on the reference target. A higher value of PRM-score indicates that the PRM-reconstructed image of the test target is less likely identical to that of the reference image, whereas PRM-score $< \delta n$ implies that they are not distinguishable. To identify the defect in the concrete structures, $2\delta n$ has been set as the threshold PRM-score. The procedure of determining the PRM-score can be followed from the example shown in figure~\ref{fig:prm}. The PRM-reconstructed image for the perfect target has been shown in figure~\ref{fig:prm}(a), and the defective target has been shown in figure~\ref{fig:prm}~(b). The PRM-score of the given example has been expressed in equation~\ref{eq:prmscore} which can be simplified to equation~\ref{eq:prm}. The PRM-score for each of the three test cases considered in the present work, have been calculated following equation~\ref{eq:prm}. The algorithm to obtain PRM-score has been described below.

\begin{equation}
\label{eq:prmscore}
\textrm{PRM-score} = \frac{\textrm{No. of pixels in `$R_{p}$' ($n$)-No. of pixels in `$R_{d}$' ($n-m$)}}{\textrm{No. of pixels in `$R_{p}$' ($n$)}}\times \frac{1}{\delta n}
\end{equation}
After simpifying,
\begin{equation}
\label{eq:prm}
\textrm{PRM-score} = \frac{m}{n\times\delta n}
\end{equation}

%\par\noindent\rule{\textwidth}{0.8pt}
\begin{flushleft}
\underline{\textbf{Algorithm of PRM-score:}}
\end{flushleft}
\begin{itemize}
    \item \textbf{Input:} Numerical matrices for 2D $S$-map of reference, $R_{p}$, and test, $R_{d}$.
    \item \textbf{Step-1:} The kernel sub-matrix `$K$', is chosen randomly from arbitrary location of the target region of $R_{p}$.
    \item \textbf{Step-2:} `$K$' is convoluted to mother matrix, `$R_{p}$' and the number of approved PRM-approved pixels is calculated ( say `$n$').
    \item \textbf{Step-3:} The random error which may arise if `Step-2' is repeated, is considered to be $\delta n$ = $\frac{1}{\sqrt n}$ according to Poisson statistics.
    \item \textbf{Step-4:} `$K$' is convoluted to test matrix, `$R_{d}$' and the number of approved PRM-approved pixels is calculated (say `$n-m$').
    Here, `$m$' is the number of pixels in the non-approved region (defective region).
    \item \textbf{Step-5:} The PRM-score is calculated using equation~\ref{eq:prm}.
    \item \textbf{Output:} The PRM-score for the test case is returned in the units of $\delta n$. Comments on similarity or difference between reference and test is provided.
\end{itemize} 

%\par\noindent\rule{\textwidth}{0.8pt}

\section{Results}
\label{sec:4}
The $S$-maps of the test cases as obtained from the Geant4 simulation have been furnished below along with PRM-reconstrcuted images. The statistical significance obtained from the t-test and the PRM-score have been listed in table~\ref{set1}.

\subsection{Detection of targets and defects}
\begin{figure}[!htbp]
	\centering
\includegraphics[trim = 0 0 0 0, clip, angle = 0, width=0.8\textwidth]{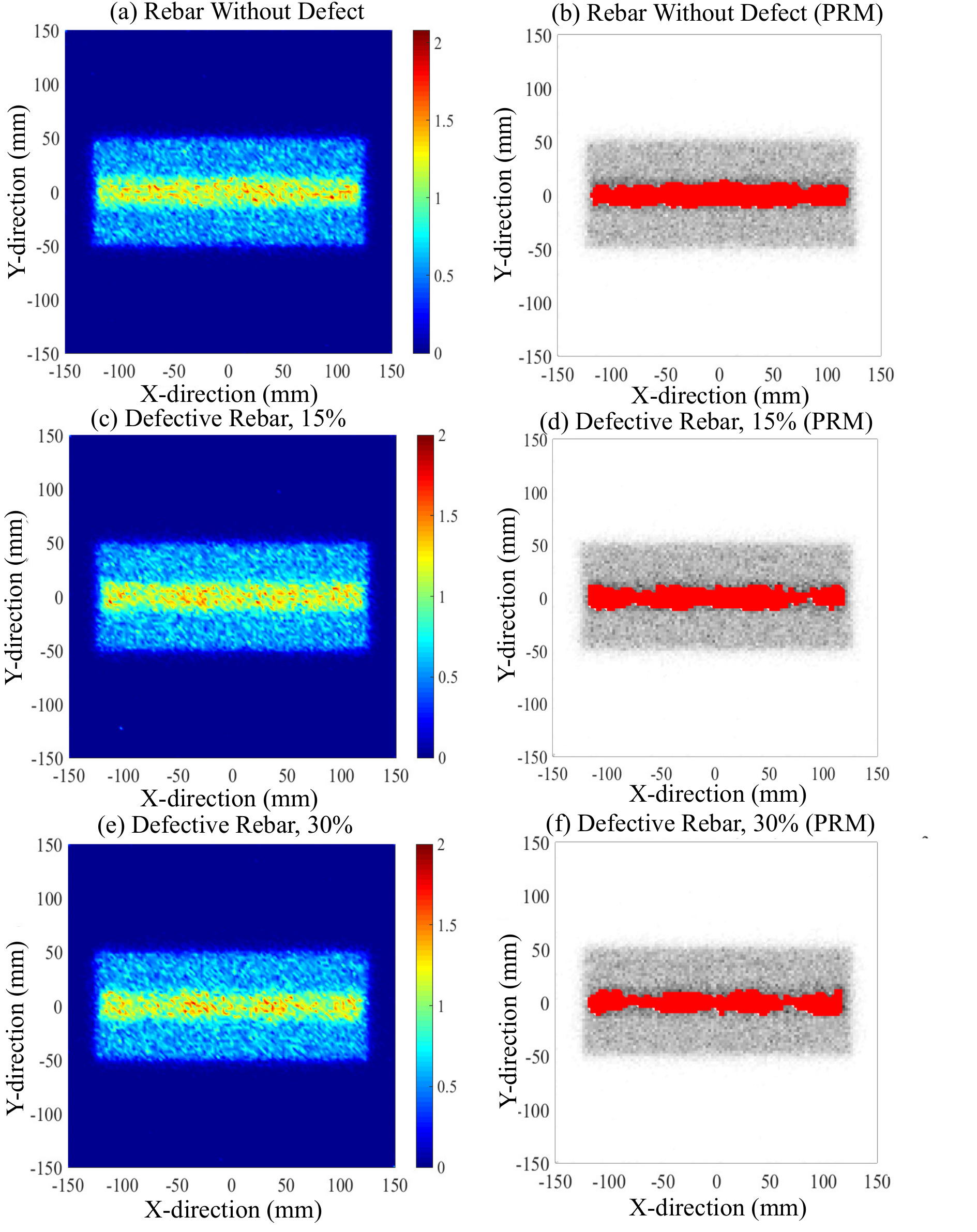}
\caption{\label{CombRebar}~(a),(b) $S$-map and PRM-reconstructed image for the perfect rebar. (c), (d) and (e), (f) same for rebar with 15\% and 30\% thicknesses rusted, respectively. The $S$-parameter has been shown in the color scale for each image.} 
\end{figure}

The figure~\ref{CombRebar} shows a collection of $S$-maps and PRM-reconstructed images of three different states of rebar (perfect, 15\% and 30\% rust). The left-hand-side plots show the $S$-maps of the three states as produced by the Geant4 simulation while the PRM-reconstructed images have been depicted on the right-hand-side. The color-axis in the $S$-maps shows the $S$-parameter accumulated in different pixels. In the PRM-reconstructed images, the pixels approved by the PRM have been shown in red while the rejected ones in grey. It can be noted from figure~\ref{CombRebar} (a) and (b) depicting the cases of perfect rebar. The concrete volume is clearly segregated from the background and the steel rebar has been discriminated from the concrete volume. In figure~\ref{CombRebar}~(c) and (e), the defective (rusted) parts can be seen with low $S$-parameter in comparison to the steel section. It can be noted that the clarity has improved with the defect percentage, \textit{i.e.}, for the case of 30\% rust, shown in figure~\ref{CombRebar} (e) and (f), the defect is much clearer than the case of 15\%. The same observation can be made from the parameters displayed in table~\ref{set1}. It can be observed from the plots in figure~\ref{CombRebar} that the PRM technique can perform reasonably well in identifying the defect although it is not very efficient in determining its shape. With 15\% defect, rust thickness becomes 2.25~mm which is smaller than the image pixel size (2.5~mm) and this sets up the lower limit of the technique. The same can be realised from the statistical significance of 2.29 which is barely above 2$\sigma$ and the PRM score 1.48 which is also less than the discrimination threshold 2$\delta n$.

\begin{figure}[!htbp]
	\centering
\includegraphics[trim = 0 0 0 0, clip, angle = 0, width=0.8\textwidth]{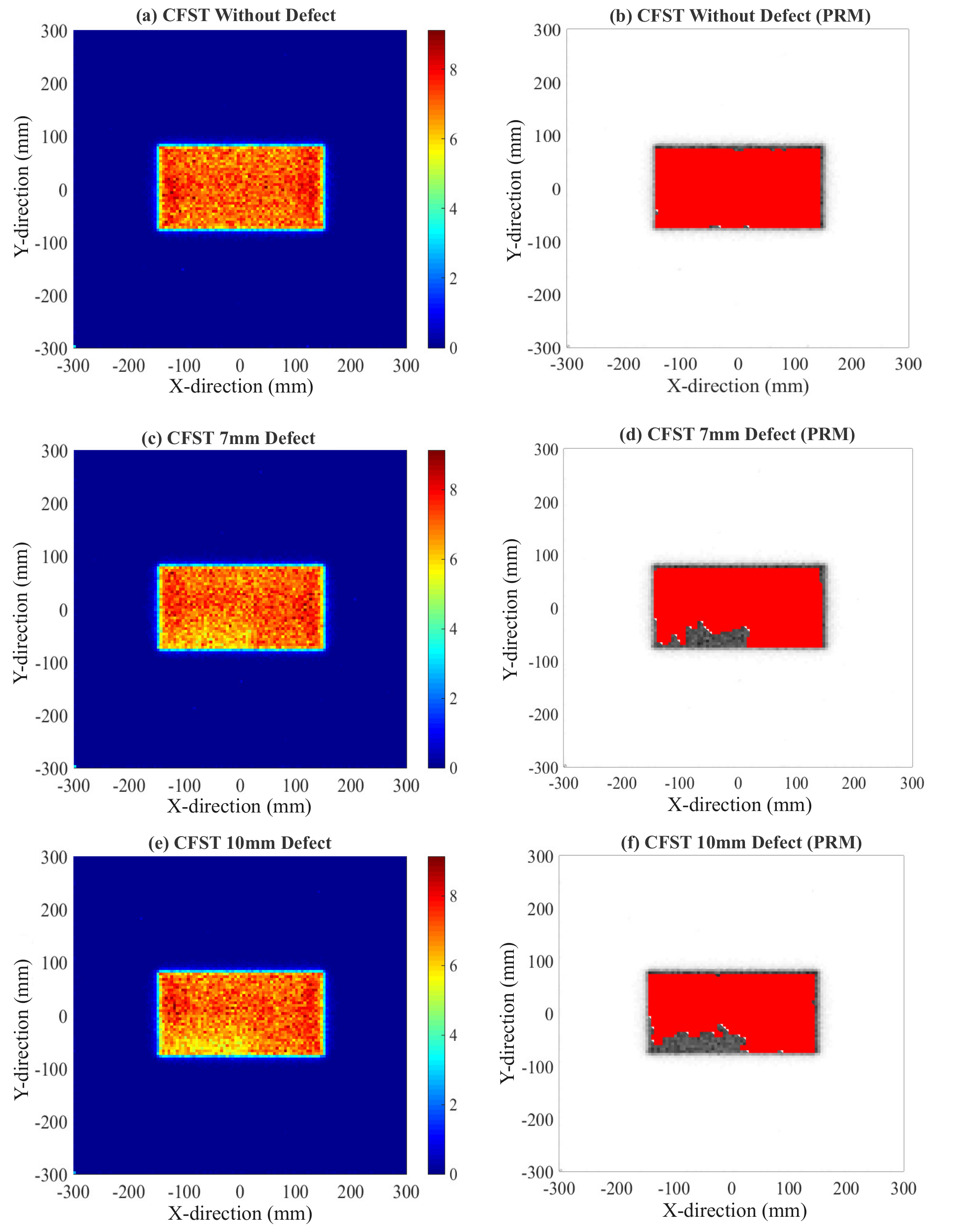}
\caption{\label{CombCFST}~(a), (b) $S$-map and PRM-reconstructed image for the CFST without defect. (c), (d) and (e), (f) same with 7~mm and 10~mm void respectively. The defects can be observed in low-$S$-parameter in $S$-maps and grey pixels in case of PRM-reconstructed images.} 
\end{figure}

Figure~\ref{CombCFST} displays the $S$-maps and corresponding PRM-reconstructed images for the perfect and defective cases of CFST. Figure~\ref{CombCFST}~(a), (b) display the CFST without any defect while images of the CFST with 7 mm and 10 mm wide circumferential voids have been shown in figure~\ref{CombCFST}~(c), (d) and (e), (f), respectively. It can be noted that the voids has been identified in both the cases of the CFST, however, without proper shape recognition. The portion of the arc-shaped void at the extreme right circumferential location (\textit{vide} $XZ$-plane view in figure~\ref{JoinCFST}) has been recognized correctly as there is no concrete. However, towards the centre the increased proportion of concrete has spoiled the identification. In addition to that, the steel covering with higher density and radiation length has contributed in the underestimation of the extent of void. It can be seen from the table~\ref{set1} that the void in both the cases has been identified with more than 3$\delta$n PRM-score and 3$\sigma$ in t-test.

\begin{figure}[!htbp]
	\centering
\includegraphics[trim = 0 0 0 0, clip, angle = 0, width=0.8\textwidth]{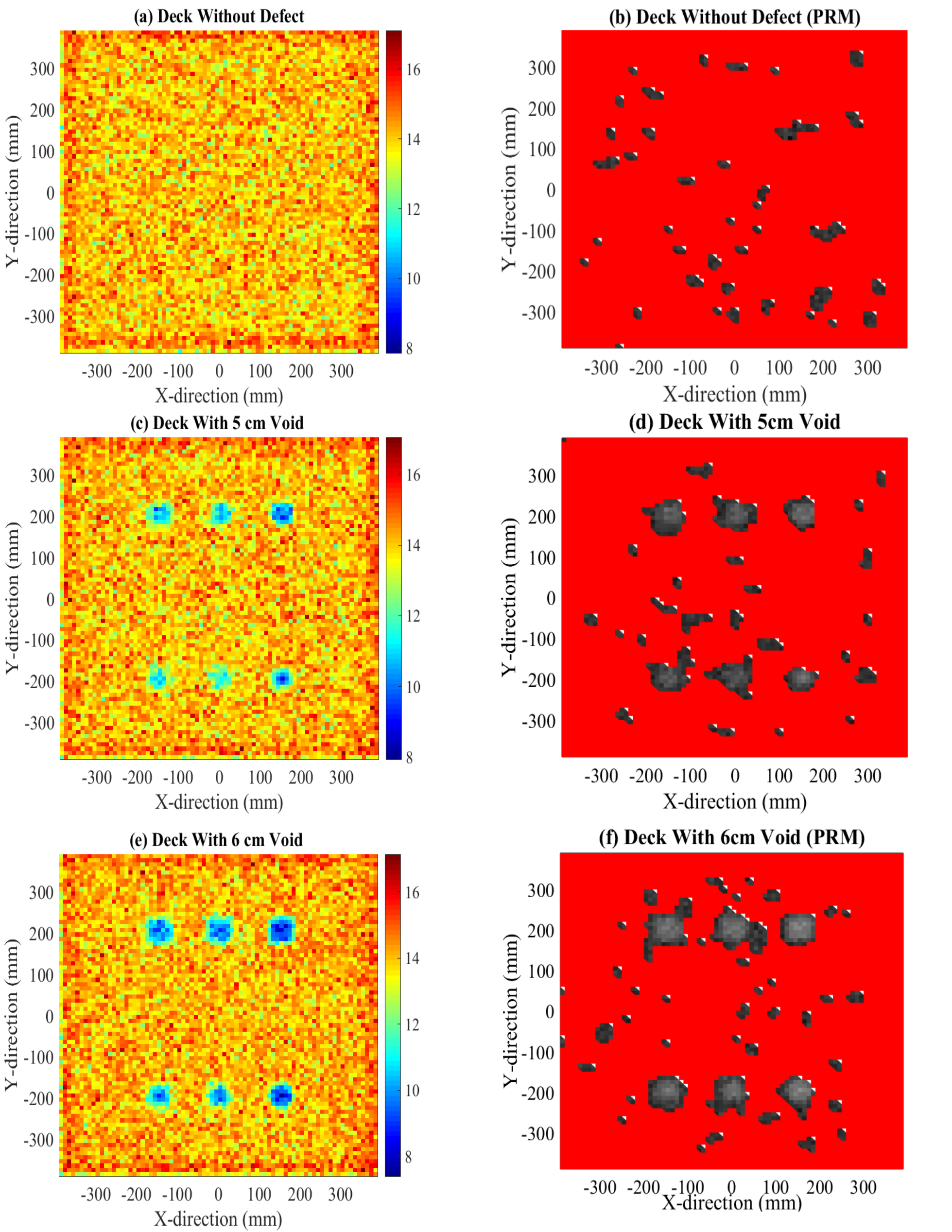}
\caption{\label{DefVoid} (a), (b) $S$-map and PRM-reconstructed image for the concrete deck without defect. (c), (d) and (e), (f) same with 5 cm and 6 cm voids, respectively. The defects can be observed in low-$S$-parameter in $S$-maps and grey pixels in case of PRM-reconstructed images. The voids at 4 cm depth are at the extreme right and depth increases from right to left in steps of 4 cm. Instead of showing the plots for the whole ROI, a magnified view has been shown for all three states of the concrete deck. This way, the voids are more noticeable.} 
\end{figure}

%Figure~\ref{DefVoid}, shows t
The $S$-maps and PRM-reconstructed images for all states of concrete deck (perfect, 33\% and 40\% defects) have been depicted in figure~\ref{DefVoid}. As this problem deals with discrimination between only two types of materials (concrete and void), this test case seems comparatively simpler than previous ones, but low-density of concrete causes makes the image reconstruction challenging. From these plots, it can be noted that the images of larger voids are clearer with better shape detection. As described in section~\ref{sec:deck}, the voids have been placed at different depths inside the concrete deck. Evidently, the sub-surface voids (at 4 cm from top) have turned out more palpable than the deep ones. Nevertheless, MST has been successful to identify the defects at different depths which is found as a deficiency of other NDE techniques, like, IR thermography, Ultrasound tomography~\cite{Abdel-Qader2008,Verma2013,Oshita2015,Dobrowolska2020}. From table~\ref{set1}, it can be noted that the concrete decks with voids have been distinguished from the no void case with more than 4$\sigma$ significance as per t-statistics and it has also crossed more than 5$\delta n$ PRM-score.

\begin{table}[h!]
  \begin{center}
    \begin{tabular}{|c|c|c|c|}
    \hline
       & & & \\
      \textbf{Target type} & \textbf{Defect dimension} & \textbf{Statistical significance} & \textbf{PRM-score}\\
       & (mm) & & \\
      \hline
      Rebar & 2.25 & 2.29  & 1.48 \\
          & 4.5 & 4.75 & 4.45 \\
      \hline
      CFST & 7 & 3.83 & 3.85  \\
          & 10 & 6.48 & 4.95 \\
     \hline
      Concrete & 50 & 4.2 & 5.62  \\
          deck & 60 & 4.89 & 7.98 \\
      \hline
    \end{tabular}
    \caption{\label{set1} The identification metrics, statistical significance of t-test and PRM-score for different defects of three test cases.}
    \end{center}
\end{table}

\begin{figure}[!htbp]
\centering
\includegraphics[trim = 0 00 0 00, clip, angle = 0, width=\textwidth]{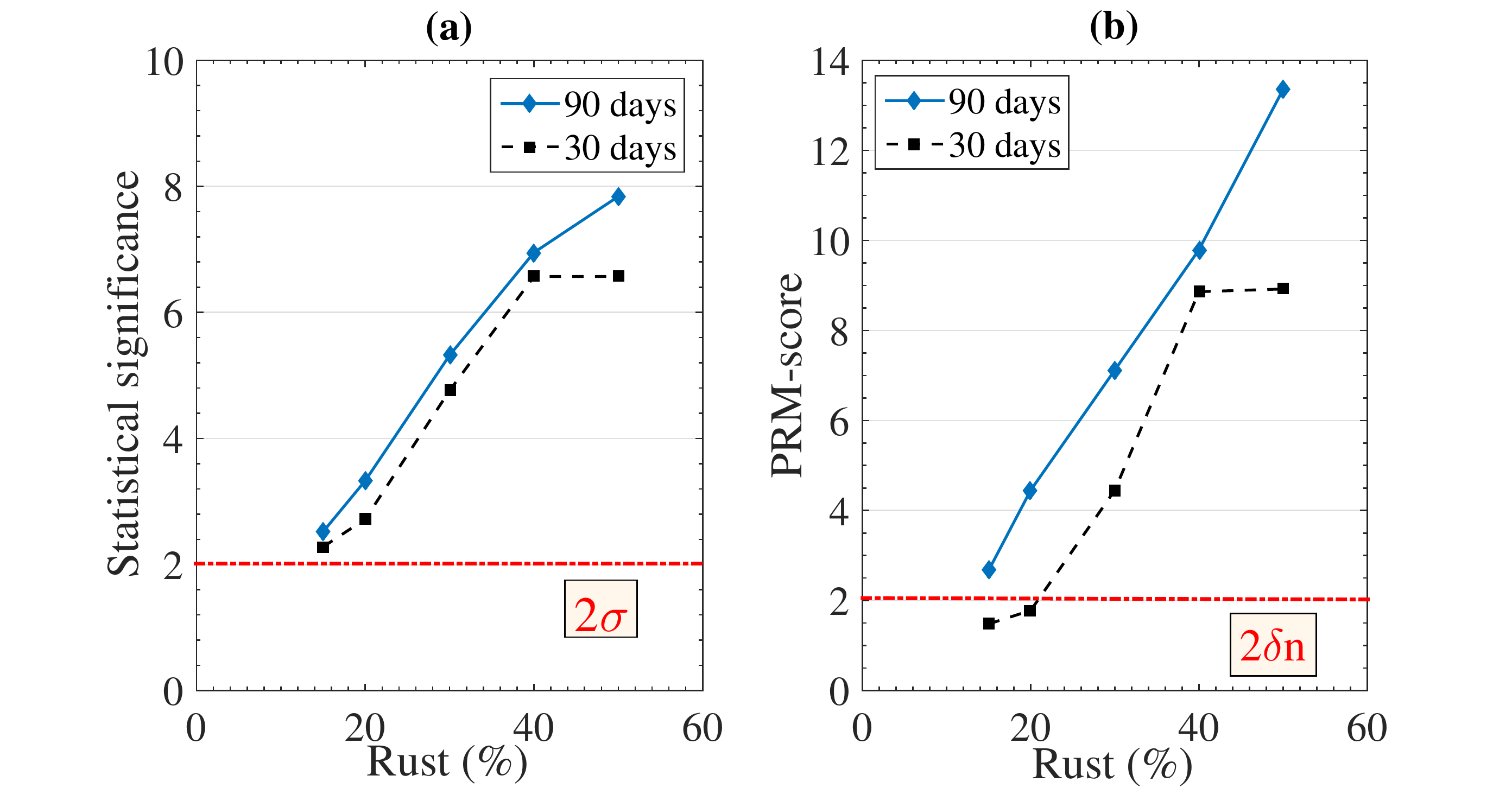}
\caption{\label{Score}~(a) The statistical significance obtained from t-statistics for different thicknesses of defect (15-50\%). (b) The PRM-score obtained for the same range of defects has been shown. The minimum threshold for identifying the defects, 2$\sigma$ for t-statistics 2$\delta n$ for PRM-score and have been marked.} 
\end{figure}

\subsection{Variation with defect thickness and muon exposure}

The limit of discrimination capability of the present MST system in imaging concrete structures has been studied with the rusted rebar in more detail. The particular test case of rebar has been chosen due to its composite geometry with three different materials, concrete, rust, and steel, which can offer a considerable variation in $Z$ and $\rho$. The study has been done on the basis of variation of defect thickness and muon exposure received by the system. The thickness of the rusted part of the rebar has been varied between 15\% to 50\%. The same set of defects have been studied for 90 days of exposure and compared to that obtained in 30 days. In each case, the defective cases have been compared to the  perfect case using t-statistics and PRM processing. The statistical significance calculated from t-test has been shown in figure~\ref{Score} (a) while the PRM-score of the present study can be found in figure~\ref{Score} (b). From the figures, it can be observed that the 15\%  and 20\% defect with PRM-score $<$ $2\delta n$ are hardly distinguishable, whereas they have been discriminated with marginally higher than $2\sigma$ significance in terms of t-statistics. The detection of defects has also improved with increase in exposure and the discrimination becomes more precise.

\section{Conclusion}
\label{sec:5}
In this work, a numerical study has been carried out to investigate the credibility of a MST system in identifying defects of civil structures. For this purpose, three test cases of steel rebar, CFST and concrete deck with some commonly-occurring defects have been considered. Each of them is unique in its structure and composition and found pretty challenging for imaging the defects using other NDE techniques. Two different dimensions of defects have been studied in each test case. This numerical study has been performed in Geant4 using an imaging setup consisting of gaseous ionization detectors as tracking devices. The identification of the defects have been done by analyzing the 2D scattering images ($S$-maps) obtained for the test cases. The $S$-maps have been processed with a pattern recognition method to improve the defect identification. The images obtained for defective cases have been compared to perfect cases and the efficacy of the MST system has been expressed in terms of statistical significance obtained from t-statistics and a metric devised in this work, namely, the PRM-score. The present MST system has been found to be able to quantify defects reasonably. It has been able to identify defects with more than $2\sigma$ precision following t-statistics in all the test cases. Similarly, defects in all the test cases have been confirmed with more than $2\delta n$ PRM-score except for the case of 15\% defect in steel rebar. Thus for the given muon exposure, the limit of detection capability found to be 2.25 mm (15\% of the thickness of rebar).
Moreover, rebars with 15\% and 30\% thickness rusted, could be distinguished from each other with more than $2\sigma$ confidence and more than $3\delta n$ PRM-score. In the same manner, de-bonding gaps of about 8\% and 12\% thickness of CFST could be distinguished from each other with more than $2\sigma$ confidence and more than $1\delta n$ PRM-score. The identification of voids of spherical and cubical shape at different depths in concrete deck have been achieved with the present MST setup. It has also been shown that with increased exposure, the precision of defect identification improves. On the whole, the authors conclude that using cosmic muons, and detectors with decent spatial resolution (200$\mu$m) and with nominal exposure (30 days), MST can be used to to identify critical defects in concrete structures.  

\section*{Acknowledgement}
The author, Sridhar Tripathy, acknowledges the support and cooperation extended by INSPIRE Division, Department of Science and Technology, Govt. of India.

%\begin{table}
%\centering
%\begin{tabular}{lll}
%\hline\noalign{\smallskip}
%\end{tabular}
% Or use
%\vspace*{5cm}  % with the correct table height
%\end{table}

%
% BibTeX users please use
\bibliographystyle{spphys85}
\bibliography{combo,library}
%\begin{thebibliography}{}
%Author, \textit{Book title} (Publisher, place year) page numbers
% etc
%\end{thebibliography}
\end{document}